\documentclass[manuscript]{emulateapj-rtx4}
\usepackage{longtable}
\usepackage{rotating, graphicx}

\newcommand{\myemail}{ipillitt@cfa.harvard.edu}
\newcommand{\xmm}{{\em XMM-Newton}}

\newcommand{\fxu}{{erg~s$^{-1}$~cm$^{-2}$}}
\newcommand{\lxu}{{erg~s$^{-1}$}}
\newcommand{\kori}{$\kappa$ Ori}

\shorttitle{The Kappa Ori cluster.}
\shortauthors{Pillitteri et al.}

\begin{document}


\title{A star forming ring around \kori\ 250 pc from the Sun.}


\author{I. Pillitteri\altaffilmark{1,2}}
\author{S. J. Wolk\altaffilmark{1}} 
\author{S. T. Megeath\altaffilmark{3}} 

\affil{Harvard-Smithsonian Center for Astrophysics, Cambridge MA 02138, USA}
\affil{INAF-Osservatorio Astronomico di Palermo, Piazza del Parlamento 1, 
90134, Palermo, Italy}
\affil{  Ritter Astrophysical Research Center, Dept. of Physics and Astronomy, 
University of Toledo, OH -- USA} 

\email{\myemail}



\begin{abstract}
X-rays are a powerful probe of activity in early stages of star formation. 
They allow us to identify young stars even after they have lost the IR signatures of circumstellar 
disks and provide constraints on their distance. Here we report on XMM-Newton observations which 
detect 121 young stellar objects (YSOs) in two fields between L1641S and \kori. 
These observations extend the Survey of Orion~A with XMM and Spitzer (SOXS). 
The YSOs are contained in a ring of gas and dust apparent at millimeter wavelengths, and in far-IR and near-IR surveys. 
The X-ray luminosity function of the young stellar objects detected in the two fields 
indicates a distance of $250-280$ pc, 
much closer than the Orion~A cloud and similar to distance estimates of \kori.  
We propose that the ring is a $5-8$ pc diameter shell that has been swept up by \kori.  
This ring contains several groups of stars detected by Spitzer and WISE including one surrounding the 
Herbig Ae/Be stars V1818~Ori. In this interpretation, the \kori\ ring is one of several shells swept 
up by massive stars within the Orion Eridanus Superbubble, and is unrelated to the 
southern portion of Orion~A / L1641 S.
\end{abstract}




\section{Introduction}
Orion OB1 is the nearest example of an actively forming OB association.  
It contains 71 O to B3 stars, two massive molecular clouds, and over 5000 young low mass stars. 
At a distance of 400-450 pc, Orion OB1 is embedded within the Orion-Eridanus superbubble, 
generated by the massive star winds and supernovae \citep{Brown1995,Bally08}. 

The southern portion of the association includes the Orion A molecular cloud with  
the massive star forming Orion Nebula Cluster (ONC, age $\sim$ 2 Myr) and the filamentary 
dark cloud complex L1641.  
 The Orion A cloud has an almost cometary shape; 
 this morphology may arise from its interaction with the older subgroups in the OB1 association.  
Unlike the ONC, the L1641 region is comprised of primarily low mass stars in small clusters, 
groups and isolation \citep{Carpenter00,Allen2008,Hsu2012, Megeath2016}.  
The southern end of the Orion A cloud (directly south of L1641 and also known as L1647 and L1648) 
appears to bifurcate and form a ring-like structure \citep{Lombardi2011}.

The Survey of Orion A with X-ray and Spitzer (SOXS) used new and archival \xmm\ observations to study 
the Orion A cloud down to a declination of $-8.5\deg$ \citep{Pillitteri2013}.  
These observations complemented existing optical and infrared (IR) observations obtained 
with {\em Spitzer} and ground facilities (\citealp{Furesz2008}, \citealp{Megeath2012}, \citealp{Hsu2012}, 
\citealp{Alves2012}, \citealp{Fang2013}).
SOXS pinpointed the population of disk-less Pre Main Sequence stars bright in X-rays and 
elusive in infrared studies. 
This population constitutes a significant fraction of the stellar content in L1641 cloud. 

The X-ray data also detected a cluster of pre-main sequence stars surrounding $\iota$~Ori which were older 
(given the paucity of stars with disks) and closer (given the brighter X-ray luminosities) than the ONC. 
This X-ray cluster appears to be the same as the foreground cluster identified by \citet{Alves2012} 
using optical observations, and may represent the oldest stellar  population 
associated with Orion~A (see \citealp{DaRio2016}).

\begin{table}[t!]
\caption{\label{targets} Properties of the \xmm\ observations. Both fields
have been observed with EPIC as primary instrument and {\em Medium} filter.}
\resizebox{\columnwidth}{!}{
\begin{tabular}{c c c c c}\\\hline
\hline
 Name   & ObsId & R.A. (J2000)  & Dec. (J2000) & Exp. Time (ks)\\\hline
Field S (KS) & 0740570201 & $5^h42^m41^s$ & $-9^d56^m45^s$ & 52.8 ($pn$) \\
Field N (KN) & 0740570101 & $5^h41^m00^s$ & $-9^d27^m00^s$ & 48.5 ($pn$) \\
\hline
\end{tabular}
}
\end{table}

We have focused our attention to the southern end of L1641 filament near the star  
\kori\ (a.k.a. Saiph).  This star is a blue supergiant B0.5 Ia  with a temperature of about 26,500 K, 
a luminosity $\sim10^5$ L$_\odot$ and strong winds with a mass loss rate of about 
$9\times10^{-7}M_\odot $ yr$^{-1}$ \citep{Crowther2006}.  
Based on PARSEC isochrones \citep{Bressan2012,Chen2014,Tang2014,Chen2015}, 
the bolometric luminosity and $T_\mathrm{eff}$ of \kori,
we estimate an age of $6.5-7$ Myr.
The parallax determined by Hipparcos was 5.04 $\pm0.22$ mas \citep{VanLeeuwen2007},  
this is a little under $200\pm10$ pc, significantly closer than the ONC. On the basis
of Ca II absorption, \citet{Megier2009} report a distance of $240\pm40$ pc, still 
significantly closer than Orion A.  

In this paper we present evidence that \kori\ is the center of a star forming ring 
significantly closer than Orion~A. 
Due to superposition, this ring is often conflated with Orion~A or even with Mon R2.  
The evidence, which will be discussed in detail below, includes:  
the Hipparcos distance to \kori, a ring of stars with disks visible in mid-IR data,   
a similar ring visible in velocity maps from CO surveys, 
and the X-ray luminosity of the YSOs in the ring
from which we infer their distances to be similar to that of \kori. 

The structure of the paper is the following: in Sect. \ref{obs} we describe new \xmm\ observations and the
data analysis, in Sect. \ref{results} we presents our results, in Sect. \ref{discussion} we discuss them
and summarize our conclusions in Sect. \ref{conclusion}.

\begin{figure*}
\begin{center}
\resizebox{0.99\textwidth}{!}{
	\includegraphics{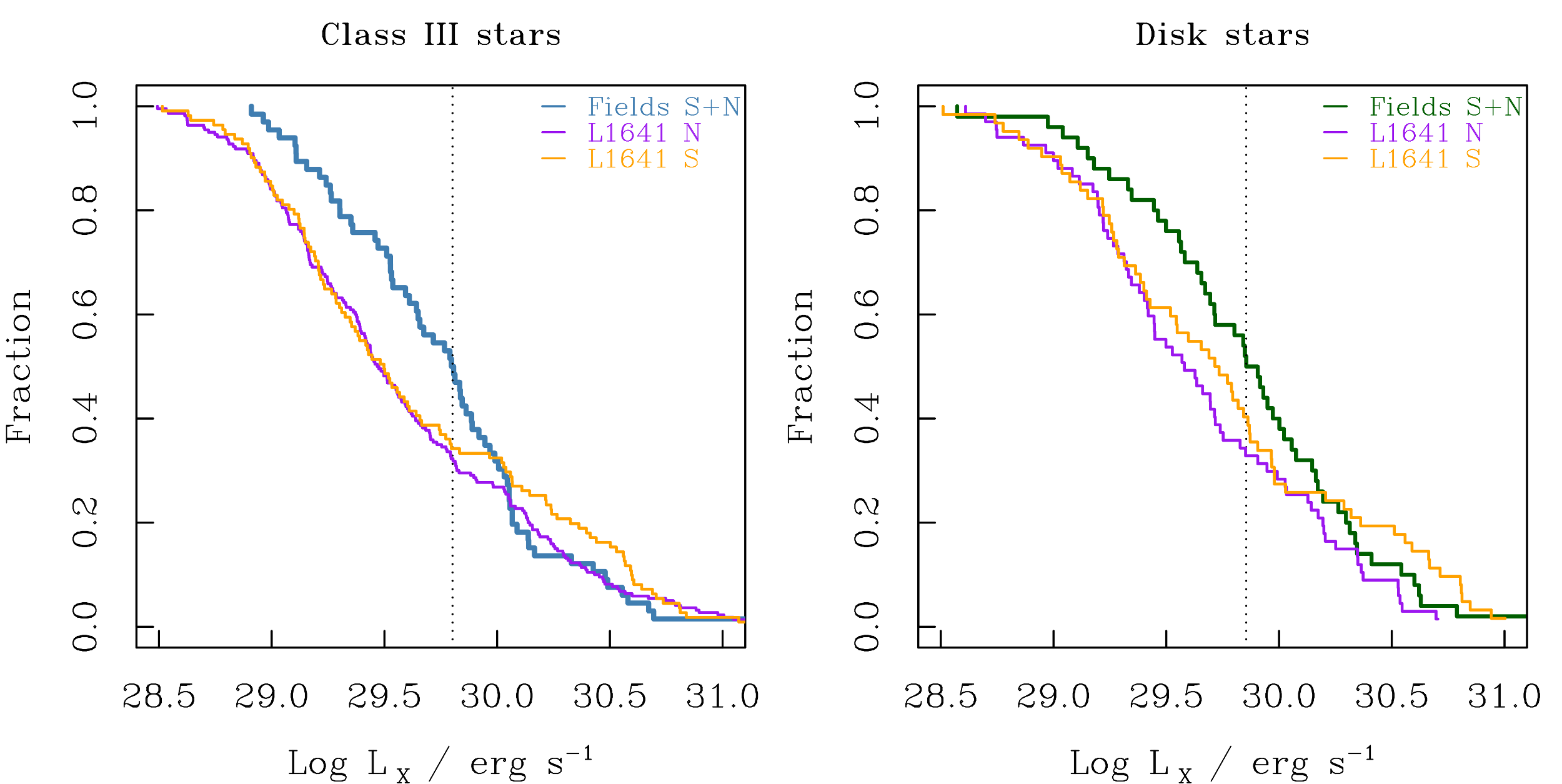} 
	}
\end{center}
\caption{\label{xlf} X-ray Luminosity distributions of  Class III stars (left panel) and stars with disks (right panel).
{ Vertical lines mark the median of the KN+KS sample. Using } the X-ray luminosities of the 
YSOs of KN and KS adopting a distance of 400 pc, these appear systematically over-luminous than 
YSOs in L1641.}
\end{figure*}

\section{X-ray observations}\label{obs}

The \xmm\ observations of two fields, North (Field N = KN) and South (Field S = KS), 
were obtained in  March 10 and March 15 2015 using EPIC as the primary instrument. 
Table \ref{targets} shows the details of the exposures, each one of duration about 50 ks, and 
taken with the {\em Medium} filter. 

We used SAS ver. 14.0 to reduce the Observation Data Files (ODFs) and to 
obtain calibrated lists of events for MOS and $pn$ instruments. 
We filtered the events in $0.3-8.0$ keV energy band, and used only events with 
$\mathrm{FLAG} = 0$ and $\mathrm{PATTERN} <12$ as prescribed by the SAS manual. 
With SAS we obtained exposure maps in $0.3-8.0$ keV and performed source detection with a code based on 
wavelet convolution that operates simultaneously on MOS and $pn$ data \citep{Damiani1997a, Damiani1997b}. 
We used a threshold of significance of 4.5 $\sigma$ of the local background to discriminate real sources 
from spurious background fluctuations. However, we added few sources to the final
list with significance in $4.0<\sigma<4.5$ for the cases of positional match with objects in 
SIMBAD or PPMX catalogs.
The final list was also checked for spurious sources that could appear at the border of the CCDs. 
{ We detected 238 X-rays sources with significance $>4\sigma$ of local background; 
104 sources are in KN and 134 in KS (Table \ref{xlist}). 
Among the output, the wavelet code returns the count rate scaled by a reference instrument. 
In Table \ref{xlist} the count rates are referred to MOS and they have been used to derive fluxes 
using PIMMS software { and adopting a {\tt WABS+APEC} model}. 
For 36 sources with more than 500 total counts we performed a best fit modeling to X-ray spectra with 
XSPEC software. 
 For faint sources we used median energies of count events and $N_H$ values 
from a distance weighted average of $N_H$ of bright sources as inputs for PIMMS.
X-ray luminosities in the $0.3-8.0$ keV band were calculated assuming first a distance of 400 pc.} 
We calculated upper limits to count rates using the same detection code for the 61 objects with IR excess 
from the catalog of \citet{Megeath2012} and not detected in X-rays. 

\section{Results}\label{results}
\subsection{IR counterparts}
We cross-correlated the positions of X-ray sources with the coordinates of the IR catalog of 
\citet{Megeath2012}.  The IR catalog is the result of a survey of Orion with {\em Spitzer} 
that produced a classification of protostars and stars with disks. 
Of the 238 X-ray sources, 191 are identified within $8\arcsec$ of one of 206 IR objects, 
99 sources in KS, 92 sources in KN. 
Some X-ray sources were multiple matches within $8\arcsec$ of IR
objects. For these cases we assigned the most likely counterparts based on IR photometry and visual inspection 
of X-rays and IR images. However, nine X-ray sources are left associated with two or three IR objects . 

Among the IR matches, we find 15 stars with disks in KN and 35 in KS with X-ray detection. 
One protostar in KN and 3 in KS are detected in X-rays.
We used X-ray detection of sources without IR excess as criteria to identify disk-less stars (hereafter Class III stars). 
We classified as Class III stars those IR objects with X-ray detections, 
with $[4.5] - [8.0] < 0.3$ mag and brighter than $[4.5] < 14$ mag. 
At the distance of ONC (400 pc), the $[4.5]\sim14$
threshold at age of $4-5$ Myr roughly identifies $M3-M4$ spectral types and masses around $0.3 M_\odot$ 
\citep{Siess2000}.
With this selection scheme we have identified 48 objects in KN and 19 in KS as Class III candidates. 

The ratio of the stars with disks and  without disks is a probe of the evolutionary stage of the stars in the
parent cloud and eventually of their ages \citep{Haisch2001,Hernandez2008}. 
The fraction of disks/total objects (among X-ray detections) is $\sim 15/(15+48) = 15/63\sim(24\pm7)\%$ in KN 
and $\sim 35/(19+35) = 35/54 \sim (65\pm14)\%$ in KS. 
This is indicative of a possible age difference between the two groups, 
with YSOs in KS being $2-3$ Myr old and those in KN about $4-5$ Myr old. 
The presence of 14 protostars and small dense groups of stars in KS are further indication of its youth \citep{Megeath2012}.  
In  comparison, L1641 shows an average ratio of disk/total objects of ($37 \pm 6$)\%, 
although the disk fraction varies through the clouds \citep{Megeath2012,Pillitteri2013}.

\subsection{X-ray luminosity}
The sensitivity of \xmm\ exposures is about $2.5\times10^{-15}$ erg s$^{-1}$ cm$^{-2}$ on axis, which corresponds to
a luminosity of $4.8\times10^{28}$ erg s$^{-1}$ 
for objects at the distance of 400 pc; X-ray luminosities of the YSOs from Megeath's catalog were calculated 
assuming this distance.
We studied how the X-ray detection rate of stars with disk vary in different strata of 
$H$ band magnitudes. We estimate completeness of $>95\%$ at $L_X \ge 3\times10^{29}$ erg s$^{-1}$
and approximately $H=12$ mag, which covers spectral types down to late K in $2-5$ Myr old YSOs. 
Below $H=13$ mag the  detection rate rapidly falls below 50\%. 
Sensitivity and completeness of the \xmm\ exposures in KS and KN are similar to that of SOXS survey in L1641.
To compare to L1641, we used only the empirical cumulative distribution of X-ray luminosities 
for both Class III and stars with disks, these curves are shown in Fig. \ref{xlf}. 
The median of the curves of stars with disks and Class III stars in KS and KN are very similar 
($\log L_X = 29.85$ and $\log L_X = 29.80$,  respectively), this is expected for very young stars.

The idea of using the XLFs to differentiate distance is owed to  \citet{Feigelson05}. 
They observed that the XLFs of the ONC, IC 348, and NGC 1333 appear similar and put forward 
the concept of an ``universal'' XLF. \citet{Wang2007} and \citet{Wang2008} found that  a similar XLF 
can be also applied to NGC 6357 and  NGC 2244. 
For ONC the XLF is well fitted by a log-normal  function with median $\log L_X = 29.3$ 
erg s$^{-1}$,  and width $\sigma \sim 1.0$ (Feigelson et al. 2005).
Over the lifetime of a cluster, the concept of a global XLF is clearly wrong since the 
XLF evolves in time and different stars evolve at different rates. 
However, for the youngest clusters differential evolutionary effects appear minimized as 
long as $L_X$ is in saturated regime ($t\le 20-30$ Myr). 
Therefore, when comparing XLFs of YSOs from \kori\ and L1641, we expect to find similar X-ray emission levels
given the similar sensitivity of \xmm\ observations.
The XLFs of KS and  KN indicate a significant systematic over-luminosity of both
stars with disks (Class II) and without disks (Class III) when compared to the analog curves of L1641, 
this difference is of order of about $0.3-0.4$ dex in the median of the distributions. 
This difference of luminosity can be explained with a wrong distance assumed for the objects in KN and KS.
We used 400 pc for the YSOs near \kori, i.e., same distance of L1641, but to match the medians of the 
distributions, the distance must be reduced by a factor $\sim\sqrt {10^{(0.3-0.4)}}\sim 1.4-1.6$,
or $d_{corrected}\sim250-280$ pc.  
This distance is consistent with the distance to \kori\ estimated from the Hipparcos parallax and 
from Ca II absorption ($240\pm40$ pc, \citealp{Megier2009}), 
and suggests that the YSOs in these two \xmm\ fields could be at 
the same distance of \kori, and significantly closer than Orion~A. 
{ For a distance 250 pc, our \xmm\ exposures have a luminosity limit 
of $L_X\sim 2.2\times10^{28}$  erg s$^{-1}$; previous  ROSAT $RASS$ survey had a
typical flux limit of $2\times10^{-13}$ erg s$^{-1}$ cm$^{-2}$, which corresponds to $1.5\times10^{30}$ \lxu. 
This limit is high enough that $RASS$ should have missed the totality of YSOs in KN and KS fields. 
At 250 pc we estimate \xmm\ to be sensitive to objects of spectral types as late as M5. }

\begin{figure*}
\begin{center}
\resizebox{1.01\textwidth}{!}{
	\includegraphics{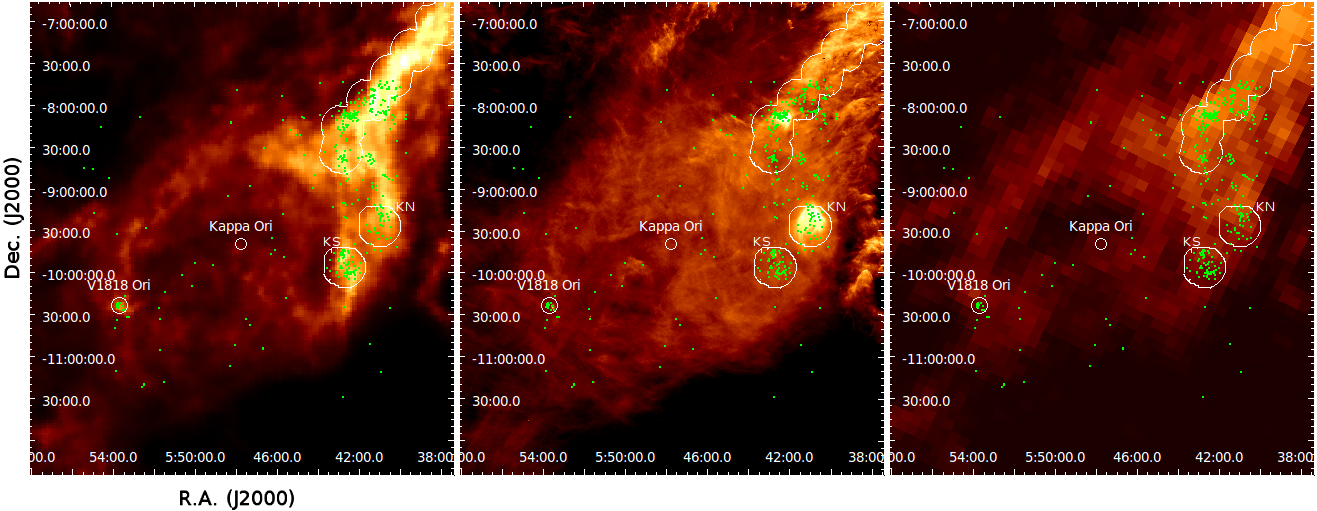} 
	}
\end{center}
\caption{\label{wisergb}
{ 
Far-IR Planck 857 GHz (left panel), mid-IR WISE $12\mu$m (central panel, image from \citealp{Meisner2014}) 
and  velocity integrated  $^{12}$CO $1-0$  image (right panel, from CfA CO survey of \citealp{Dame2001}) 
around \kori. Green symbols mark the positions of WISE objects with IR excess. 
Contours are the footprints of XMM-Newton observations. The dust ring is visible at far-IR and millimeter wavelengths 
while it appears like a bubble of diffuse emission in mid IR. 
}}
\end{figure*}

\section{A shell of stars associated with \kori.}
\label{discussion}
{
The YSOs found by \xmm\ are coincident in a structure of dust and gas shaped like a bubble
in mid-IR (Fig. \ref{wisergb}), and apparent as a ring in near-IR extinction maps \citep{Lombardi2011}, in far-IR dust emission 
maps \citep{Lombardi2014}, and CO maps (see Fig. 5a in \citealp{Wilson2005}. 
We propose that this ring is tracing a partial, inhomogeneous shell centered on \kori.  
The entire shell shows strong mid-IR emission in WISE $12\mu$m, suggesting that the inner surfaces of the shell are heated by 
\kori. Although the $12\mu$m emission is more uniform than at longer wave-lengths, structures apparent in Planck and CO images 
are also visible at $12\mu$m, bolstering confidence that cold dust and gas structures are related. 
At the distance of \kori\ ($\sim$250 pc), the radius of the bubble is $\sim8$ pc. 
The heated walls of a small cavity is also visible in the WISE $12\mu$m data toward KN, and due to an F0 star 
within the dust ring. Another similar cavity is visible in L1641 S, this other one is likely part of Orion A.
}

\begin{figure*}
\begin{center}
\resizebox{0.99\textwidth}{!}{
	\includegraphics{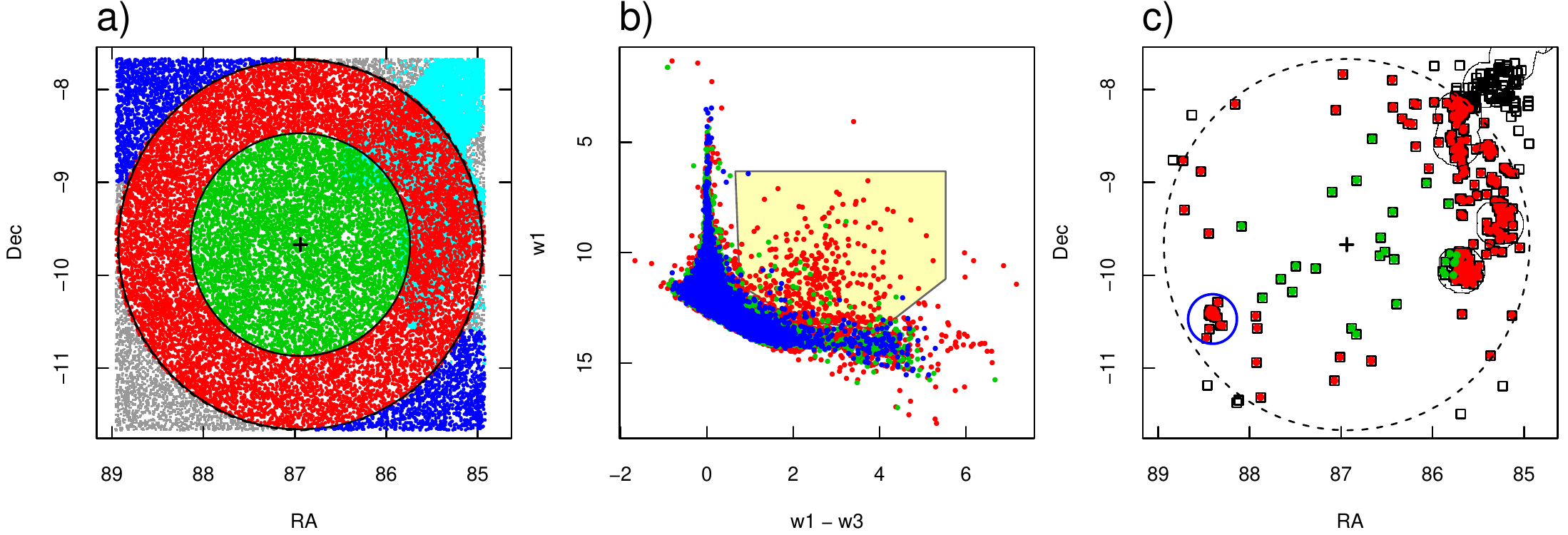} 
	}
\end{center}
\caption{\label{wiseplots} 
a)  Map of WISE objects around \kori. Red: objects in a ring $1.2-2$ deg from \kori;
green: objects within 1.2 $\deg$  from \kori; blue: control sample ($d>2\deg$); cyan: catalog of \citet{Megeath2012}. 
The control fields shown by the blue dots extend to the NE and SW beyond the limits of the display plot, for a 
coverage of 3.7 square degrees.
b)  $w1$ vs. $w1-w3$ for objects in the ring and the control sample. The yellow area delimit where
most of YSOs with IR excesses are found.  
c) Map of the objects with IR excesses, symbol colors as in a). Most of these objects fall within the circle
of $<2 \deg$ from \kori. Clumps of YSOs are recognized both toward KN\&KS and the southern end of L1641 
and toward V1818~Ori (marked with a blue circle). }
\end{figure*}

Fig. \ref{wiseplots} shows a map of the position of WISE objects in a
box $2\times2$ sq. deg around \kori. 
Objects within 1.2 deg from \kori\ and in the ring $1.2-2$ deg from \kori\ are marked with green and red colors,
respectively. Blue and cyan objects are the control samples and the catalog of \citet{Megeath2012}, respectively. 
Central panel of Fig. \ref{wiseplots} shows a color-magnitude diagram (CMD) of the same objects 
using $w1$ and $w3$ WISE band.  A significant fraction of objects in the ring show excess in $w1-w3$ color. 
These stars show similar colors and magnitudes of the stars with disks of the Megeath's catalog. 
The region of objects with IR excesses is marked with the yellow area in the CMD.
The density of IR-excess sources in the control field is similar to that in the green region, hence
we find no clear evidence of IR-excess sources inside the ring.
Across the ring the surface density of IR-excess objects is four times larger than that found within 1.2 deg 
from \kori, indicating a population of young stars embedded in the ring.

\subsection{V1818~Ori}
The V1818 Ori’s clump sits on the same ring but eastward of \kori, 
on the opposite side with respect to KN/KS fields (see Fig. \ref{wisergb}). 
V1818~Ori is a Herbig Be object of spectral type B7. 
\citet{Chiang2015} has studied V1818~Ori and its spectrum, and identified two dozens of nearby YSOs 
that appear to form a small group together with the Herbig Be star.  
They have associated V1818~Ori and the nearby YSOs to the Mon~R2 complex at $\sim900$ pc 
based on the similarity of radial velocity of V1818~Ori with the velocities of the CO map of the 
poorly studied nebula of NGC~2149 and to Mon~R2 association. 
The value of extinction in near-IR bands and the $J$ band magnitude of V1818~Ori are fully consistent 
with an object of spectral type B7 V star at 250 pc. 
Were it at 900 pc as hypothesized by \citet{Chiang2015},
V1818~Ori should have the luminosity of a more evolved B7 III star suffering of a much higher extinction.

Given the location of this group of YSOs on the dust ring, we speculate that they are 
part of the same \kori\ cluster at $250-280$ pc. 
The velocity of V1818~Ori is similar to velocity components of the ring observed in the CO 
map of \citet{Wilson2005} adding further evidence of its association to \kori.

\subsection{The \kori\ star forming shell.}
We propose that the YSOs in KN and KS fields and in the V1818 Ori region are
associated with a broken shell of dense gas surrounding \kori.  Furthermore, we present evidence that this shell, 
although it overlaps the  L1641 cloud in projection, is closer than the Orion~A at a distance of $250-280$ pc. 
The shell is detected in near-IR extinction, far-IR dust continuum and CO maps indicating that it has molecular gas. 
Some of the YSOs found in the \kori\ region by \citet{Megeath2012} may belong to the shell; a bimodal distribution 
of the X-ray luminosities in L1641 S suggest that a fraction of the stars in L1641~S may be part of this foreground cluster. 
The WISE survey shows no clear evidence for a population of young stars with disks interior to the shell beyond 
the estimated rate of contamination; however, given the age estimates for \kori\ of 7 Myr, the dusty gas disks 
should have disappeared and would not be detectable by WISE.

\citet{Ochsendorf2015} present a revised picture of the Orion-Eridanus super bubble, establishing a larger size 
of the super bubble and demonstrating that Barnard’s Loop, the $\lambda$~Ori bubble, as well as G206-17+13 
are shells powered by supernovae and stellar winds expanding into the super bubble.  
In cases where the shells are not ionized by the OB association, such as in the $\lambda$~Ori bubble, 
the gas can cool and collapse to form the next generation of stars. 
The presence of intermediate velocity absorption features toward \kori\ confirms that is located in 
the super bubble.   

While the estimated age of \kori\ is 7 Myr, two velocity components in the 
CO ($J=1-0$) lines separated by 7 km s$^{-1}$ imply a dynamical age of the shell of 1.5 Myr 
\citep{Wilson2005}. This means that there was sufficient time for the winds of \kori\ to sweep up the shell.  
The star formation in the shell is $2-5$ Myr old, consistent with a scenario in which the formation of 
these stars has been triggered by \kori.  

\section{Conclusions}
\label{conclusion}
We have presented the results from two \xmm\ observations of two regions west to \kori\ containing a few dozens of 
YSOs. We have detected a total of 238 X-ray sources, 134 sources in field KS and 104 in field KN. 
We found matches in the IR catalog of YSOs by \citet{Megeath2012} for 206 of them, 
50 X-ray sources are stars with disks, 4 protostars and 57 candidates Class III stars. 

We find evidence of a number of YSOs surrounding \kori\ and completely unrelated to Orion~A.
The distributions of X-ray luminosities show that the YSOs are incompatible with the distance of Orion~A
and ONC. Rather, a closer distance of $250-280$ pc is favored, the same distance as measured to \kori. 
Together with evidence from mid-IR and far-IR photometry and CO maps, we speculate
that the YSOs in the \xmm\ fields, as well the group of YSOs around V1818~Ori
and some YSOs in the very southern tail of L1641, form a distinct star forming shell with \kori\ 
as its most massive member. 

If confirmed by future measurements, in particular those of GAIA, 
the \kori\ ring will be one of several nested shells found in the Orion-Eridanus superbubble 
as examples of star formation mechanisms where massive stars sweep up the gas and dust 
in the hot interiors of superbubbles to form shells of cold, star forming gas.

\acknowledgments
S.T.M. acknowledges useful discussions with Noel Richardson and Adolf Witt.
S.J.W. was supported by NASA contract NAS8-03060.

{\it Facilities:} \facility{XMM-Newton, WISE, Planck}.

\appendix

\begin{table}[ht]
\centering
\caption{List of X-ray sources detected in fields KN and KS. } 
\label{xlist}
\resizebox{0.95\textwidth}{!}{
\begin{tabular}{rccrrrcccrr}
  \hline
  \hline
 N & $\alpha$ & $\delta$ & Rate & Rate error & Flux & Proto & Disks & C3 & [3.6] & [3.6]-[12] \\ 
   & J2000 & J2000 &  ct ks$^{-1}$ & ct ks$^{-1}$ & $10^{-14}$\fxu\ & & & & mag & mag \\ 
  \hline
  1 & 05 42 26.0 & -10 09 02.3 & 0.9 & 0.2 & 3.06 & 0 & 0 & 1 & 17.04 & 4.5 \\ 
  2 & 05 42 56.8 & -10 08 47.5 & 2.5 & 0.9 & 7.01 &  &  &  & 17.12 & 4.56 \\ 
  3 & 05 42 54.0 & -10 08 43.2 & 1.6 & 0.4 & 3.71 & 0 & 0 & 1 &  &  \\ 
  4 & 05 43 11.3 & -10 08 34.3 & 1.1 & 0.3 & 3.66 & 0 & 0 & 0 &  &  \\ 
  5 & 05 42 26.8 & -10 08 20.1 & 0.7 & 0.2 & 1.83 & 0 & 0 & 0 & 17.23 & 4.66 \\ 
  6 & 05 42 13.9 & -10 07 20.7 & 1.1 & 0.5 & 1.6 &  &  &  &  &  \\ 
  7 & 05 42 34.5 & -10 06 54.6 & 0.8 & 0.2 & 1.47 & 0 & 0 & 1 &  &  \\ 
  8 & 05 42 28.5 & -10 06 33.6 & 0.4 & 0.1 & 0.93 & 0 & 1 & 0 & 17.21 & 4.59 \\ 
  9 & 05 42 17.7 & -10 06 09.8 & 1.9 & 0.3 & 2.78 & 0 & 0 & 1 &  &  \\ 
 10 & 05 42 06.3 & -10 05 58.1 & 0.5 & 0.2 & 0.8 & 0 & 1 & 0 &  &  \\ 
 11 & 05 42 33.4 & -10 05 32.1 & 1.9 & 0.3 & 3.39 & 0 & 0 & 1 & 12.34 & 0.14 \\ 
 12 & 05 41 59.0 & -10 05 16.7 & 0.4 & 0.3 & 0.59 & 0 & 0 & 1 & 17.25 & 4.52 \\ 
 13 & 05 42 55.7 & -10 04 50.4 & 1.1 & 0.2 & 1.61 & 0 & 1 & 0 &  &  \\ 
 14 & 05 42 30.0 & -10 04 37.4 & 0.5 & 0.1 & 0.82 & 0 & 0 & 1 & 17.48 & 4.7 \\ 
 15 & 05 42 51.8 & -10 04 32.1 & 0.6 & 0.1 & 0.98 & 0 & 0 & 0 &  &  \\ 
\hline
\end{tabular}
}
\tablecomments{Table \ref{xlist} is published in its entirety in the electronic edition of
the journal.  A portion is shown here for guidance regarding its
form and content. Unabsorbed fluxes are given in 0.3-8.0 keV band.
Proto, Disks and C3 are flags for the classification as protostars, stars with disks
\citep{Megeath2012} or Class III stars. For sources with WISE counterparts,  [3.6] magnitudes 
and [3.6] -[12] color are listed.}

\end{table}

\begin{thebibliography}{}
\expandafter\ifx\csname natexlab\endcsname\relax\def\natexlab#1{#1}\fi

\bibitem[{{Allen} \& {Davis}(2008)}]{Allen2008}
{Allen}, L.~E., \& {Davis}, C.~J. 2008, {Low Mass Star Formation in the Lynds
  1641 Molecular Cloud}, ed. {Reipurth, B.}, 621

\bibitem[{{Alves} \& {Bouy}(2012)}]{Alves2012}
{Alves}, J., \& {Bouy}, H. 2012, \aap, 547, A97

\bibitem[{{Bally}(2008)}]{Bally08}
{Bally}, J. 2008, {Overview of the Orion Complex} (Reipurth, B.), 459--+

\bibitem[{{Bressan} {et~al.}(2012){Bressan}, {Marigo}, {Girardi}, {Salasnich},
  {Dal Cero}, {Rubele}, \& {Nanni}}]{Bressan2012}
{Bressan}, A., {Marigo}, P., {Girardi}, L., {et~al.} 2012, \mnras, 427, 127

\bibitem[{{Brown} {et~al.}(1995){Brown}, {Hartmann}, \& {Burton}}]{Brown1995}
{Brown}, A.~G.~A., {Hartmann}, D., \& {Burton}, W.~B. 1995, \aap, 300, 903

\bibitem[{{Carpenter}(2000)}]{Carpenter00}
{Carpenter}, J.~M. 2000, \aj, 120, 3139

\bibitem[{{Chen} {et~al.}(2015){Chen}, {Bressan}, {Girardi}, {Marigo}, {Kong},
  \& {Lanza}}]{Chen2015}
{Chen}, Y., {Bressan}, A., {Girardi}, L., {et~al.} 2015, \mnras, 452, 1068

\bibitem[{{Chen} {et~al.}(2014){Chen}, {Girardi}, {Bressan}, {Marigo},
  {Barbieri}, \& {Kong}}]{Chen2014}
{Chen}, Y., {Girardi}, L., {Bressan}, A., {et~al.} 2014, \mnras, 444, 2525

\bibitem[{{Chiang} {et~al.}(2015){Chiang}, {Reipurth}, \&
  {Hillenbrand}}]{Chiang2015}
{Chiang}, H.-F., {Reipurth}, B., \& {Hillenbrand}, L. 2015, \aj, 149, 108

\bibitem[{{Crowther} {et~al.}(2006){Crowther}, {Lennon}, \&
  {Walborn}}]{Crowther2006}
{Crowther}, P.~A., {Lennon}, D.~J., \& {Walborn}, N.~R. 2006, \aap, 446, 279

\bibitem[{{Da Rio} {et~al.}(2016){Da Rio}, {Tan}, {Covey}, {Cottaar}, {Foster},
  {Cullen}, {Tobin}, {Kim}, {Meyer}, {Nidever}, {Stassun}, {Chojnowski},
  {Flaherty}, {Majewski}, {Skrutskie}, {Zasowski}, \& {Pan}}]{DaRio2016}
{Da Rio}, N., {Tan}, J.~C., {Covey}, K.~R., {et~al.} 2016, \apj, 818, 59

\bibitem[{{Dame} {et~al.}(2001){Dame}, {Hartmann}, \& {Thaddeus}}]{Dame2001}
{Dame}, T.~M., {Hartmann}, D., \& {Thaddeus}, P. 2001, \apj, 547, 792

\bibitem[{{Damiani} {et~al.}(1997{\natexlab{a}}){Damiani}, {Maggio}, {Micela},
  \& {Sciortino}}]{Damiani1997b}
{Damiani}, F., {Maggio}, A., {Micela}, G., \& {Sciortino}, S.
  1997{\natexlab{a}}, \apj, 483, 350

\bibitem[{{Damiani} {et~al.}(1997{\natexlab{b}}){Damiani}, {Maggio}, {Micela},
  \& {Sciortino}}]{Damiani1997a}
---. 1997{\natexlab{b}}, \apj, 483, 370

\bibitem[{{Fang} {et~al.}(2013){Fang}, {Kim}, {van Boekel}, {Sicilia-Aguilar},
  {Henning}, \& {Flaherty}}]{Fang2013}
{Fang}, M., {Kim}, J.~S., {van Boekel}, R., {et~al.} 2013, \apjs, 207, 5

\bibitem[{{Feigelson} {et~al.}(2005){Feigelson}, {Getman}, {Townsley},
  {Garmire}, {Preibisch}, {Grosso}, {Montmerle}, {Muench}, \&
  {McCaughrean}}]{Feigelson05}
{Feigelson}, E.~D., {Getman}, K., {Townsley}, L., {et~al.} 2005, \apjs, 160,
  379

\bibitem[{{F{\H u}r{\'e}sz} {et~al.}(2008){F{\H u}r{\'e}sz}, {Hartmann},
  {Megeath}, {Szentgyorgyi}, \& {Hamden}}]{Furesz2008}
{F{\H u}r{\'e}sz}, G., {Hartmann}, L.~W., {Megeath}, S.~T., {Szentgyorgyi},
  A.~H., \& {Hamden}, E.~T. 2008, \apj, 676, 1109

\bibitem[{{Haisch} {et~al.}(2001){Haisch}, {Lada}, \& {Lada}}]{Haisch2001}
{Haisch}, Jr., K.~E., {Lada}, E.~A., \& {Lada}, C.~J. 2001, \apjl, 553, L153

\bibitem[{{Hern{\'a}ndez} {et~al.}(2008){Hern{\'a}ndez}, {Hartmann}, {Calvet},
  {Jeffries}, {Gutermuth}, {Muzerolle}, \& {Stauffer}}]{Hernandez2008}
{Hern{\'a}ndez}, J., {Hartmann}, L., {Calvet}, N., {et~al.} 2008, \apj, 686,
  1195

\bibitem[{{Hsu} {et~al.}(2012){Hsu}, {Hartmann}, {Allen}, {Hern{\'a}ndez},
  {Megeath}, {Mosby}, {Tobin}, \& {Espaillat}}]{Hsu2012}
{Hsu}, W.-H., {Hartmann}, L., {Allen}, L., {et~al.} 2012, \apj, 752, 59

\bibitem[{{Lombardi} {et~al.}(2011){Lombardi}, {Alves}, \&
  {Lada}}]{Lombardi2011}
{Lombardi}, M., {Alves}, J., \& {Lada}, C.~J. 2011, \aap, 535, A16

\bibitem[{{Lombardi} {et~al.}(2014){Lombardi}, {Bouy}, {Alves}, \&
  {Lada}}]{Lombardi2014}
{Lombardi}, M., {Bouy}, H., {Alves}, J., \& {Lada}, C.~J. 2014, \aap, 566, A45

\bibitem[{{Megeath} {et~al.}(2012){Megeath}, {Gutermuth}, {Muzerolle},
  {Kryukova}, {Flaherty}, {Hora}, {Allen}, {Hartmann}, {Myers}, {Pipher},
  {Stauffer}, {Young}, \& {Fazio}}]{Megeath2012}
{Megeath}, S.~T., {Gutermuth}, R., {Muzerolle}, J., {et~al.} 2012, \aj, 144,
  192

\bibitem[{{Megeath} {et~al.}(2016){Megeath}, {Gutermuth}, {Muzerolle},
  {Kryukova}, {Hora}, {Allen}, {Flaherty}, {Hartmann}, {Myers}, {Pipher},
  {Stauffer}, {Young}, \& {Fazio}}]{Megeath2016}
---. 2016, \aj, 151, 5

\bibitem[{{Megier} {et~al.}(2009){Megier}, {Strobel}, {Galazutdinov}, \&
  {Kre{\l}owski}}]{Megier2009}
{Megier}, A., {Strobel}, A., {Galazutdinov}, G.~A., \& {Kre{\l}owski}, J. 2009,
  \aap, 507, 833

\bibitem[{Meisner \& Finkbeiner(2014)}]{Meisner2014}
Meisner, A.~M., \& Finkbeiner, D.~P. 2014, The Astrophysical Journal, 781, 5

\bibitem[{{Ochsendorf} {et~al.}(2015){Ochsendorf}, {Brown}, {Bally}, \&
  {Tielens}}]{Ochsendorf2015}
{Ochsendorf}, B.~B., {Brown}, A.~G.~A., {Bally}, J., \& {Tielens}, A.~G.~G.~M.
  2015, \apj, 808, 111

\bibitem[{{Pillitteri} {et~al.}(2013){Pillitteri}, {Wolk}, {Megeath}, {Allen},
  {Bally}, {Gagn{\'e}}, {Gutermuth}, {Hartman}, {Micela}, {Myers}, {Oliveira},
  {Sciortino}, {Walter}, {Rebull}, \& {Stauffer}}]{Pillitteri2013}
{Pillitteri}, I., {Wolk}, S.~J., {Megeath}, S.~T., {et~al.} 2013, \apj, 768, 99

\bibitem[{{Siess} {et~al.}(2000){Siess}, {Dufour}, \& {Forestini}}]{Siess2000}
{Siess}, L., {Dufour}, E., \& {Forestini}, M. 2000, \aap, 358, 593

\bibitem[{{Tang} {et~al.}(2014){Tang}, {Bressan}, {Rosenfield}, {Slemer},
  {Marigo}, {Girardi}, \& {Bianchi}}]{Tang2014}
{Tang}, J., {Bressan}, A., {Rosenfield}, P., {et~al.} 2014, \mnras, 445, 4287

\bibitem[{{van Leeuwen}(2007)}]{VanLeeuwen2007}
{van Leeuwen}, F. 2007, \aap, 474, 653

\bibitem[{{Wang}(2007)}]{Wang2007}
{Wang}, J. 2007, PhD thesis, The Pennsylvania State University

\bibitem[{{Wang} {et~al.}(2008){Wang}, {Townsley}, {Feigelson}, {Broos},
  {Getman}, {Rom{\'a}n-Z{\'u}{\~n}iga}, \& {Lada}}]{Wang2008}
{Wang}, J., {Townsley}, L.~K., {Feigelson}, E.~D., {et~al.} 2008, \apj, 675,
  464

\bibitem[{{Wilson} {et~al.}(2005){Wilson}, {Dame}, {Masheder}, \&
  {Thaddeus}}]{Wilson2005}
{Wilson}, B.~A., {Dame}, T.~M., {Masheder}, M.~R.~W., \& {Thaddeus}, P. 2005,
  \aap, 430, 523

\end{thebibliography}
\end{document}